\documentclass[aps,floats]{revtex4}

\usepackage{epsfig}
\usepackage{graphics}

\renewcommand{\H} {{\cal H}}

\def\oppropto{\mathop{\propto}} 
\def\opsim{\mathop{\sim}}

\begin{document}

\title{Distribution of time scales in the Sherrington-Kirkpatrick
  model} \author{Alain Billoire} \affiliation{ Institut de physique
  th\'{e}orique, CEA Saclay and CNRS, 91191 Gif-sur-Yvette, France}
\begin{abstract}
Numerical data on the probability distribution of the equilibrium
relaxation time of the Sherrington-Kirkpatrick model are obtained by
means of dynamical Monte Carlo simulation, for several values of the
system size $N$ and temperature $T$.  Proper care is taken that the
thermal fluctuations on the relaxation time estimates are totally
negligible compared to the disorder induced fluctuations.  The
probability distribution of $\ln\tau-\overline{\ln\tau}$ scales with
the scaling variable $N^{1/3} (T_c-T)$ strengthening the belief that
$\overline{\ln\tau}\propto N^{1/3}$ in the whole spin glass phase.
\end{abstract}

\maketitle

The equilibrium dynamics of the Sherrington-Kirkpatrick (SK) model
remains a subject of much interest. The standard picture of the spin
glass phase of this model is that of a complex hierarchical free energy
landscape, with many valleys that correspond to pure or metastable
states. In the thermodynamic limit, both the number of valleys and the
height of the typical free energy barrier between two valleys go to
infinity. Accordingly the relaxation time of the system diverges as
the number of spins $N$ goes to infinity.

The behavior of the equilibrium relaxation time of the model $\tau$
with the system size $N$ has been studied by analytical
methods~\cite{RODMOO,VERVIR,KIN}, direct Monte Carlo simulation of the
Metropolis dynamics of the model~\cite{MACYOU,{COLBOR},BIMA}, and
indirect determination of the largest barrier
height~\cite{JANKE}. There are reasonable indications that below $T_c$
the disorder averaged relaxation time behaves according to
$\overline{\ln{\tau}} \sim \overline{B} /T \propto N^{\psi}$ as
$N\to\infty$, with $B$ the largest barrier height and an exponent
$\psi \simeq 1/3$, for both binary and Gaussian distributions of the
couplings. The behavior of the relaxation time of this model with the
system size $N$ has also been studied in the aging (non equilibrium)
regime~\cite{Hajime1,Hajime2}.

A different numerical approach to this problem has been used recently
by Monthus and Garel~\cite{MONGAR}.  (This method has already been
used in~\cite{Nightingale} for the 2d Ising model and
in~\cite{Hajime1} for the SK model.)  They use the well know mapping
(see e.g. chapter 4 of~\cite{Zinn}) of the master equation for the
Monte Carlo dynamics onto an Schr\"odinger equation in configuration
space with some quantum Hamiltonian $\H_J$ (where $J$ stands for a
particular disorder configuration).  The ground state of $\H_J$ has
zero energy, corresponding to the equilibrium stationary state of the
Monte Carlo dynamics.  The next eigenvalue is the inverse of the
largest relaxation time $\tau_J$ of the Monte Carlo dynamics of the
original model (the so called exponential relaxation time of the
dynamics~\cite{SOK}). The problem of determining $\tau_J$ is thus
reduced to the problem of finding the lowest eigenvalues of a real
symmetric sparse matrix (of size $2^N \times 2^N$), which can be
obtained with high accuracy using a standard computer routine. The
process has to be repeated for a large number of disorder
configurations $J$.

Compared to the direct Monte Carlo method, the method of Monthus and
Garel has two clear advantages: i) It is not affected by thermal
noise; ii) the long tail of the probability distribution
$P_N(\ln{\tau})$ can be easily sampled.  Indeed, provided a good
starting point is guessed, the convergence of the eigenvalue search is
very fast. The method is however limited to very small system sizes,
indeed the analysis of~\cite{MONGAR} relies on systems with $6\leq N
\leq 20$, to be compared with the nine years old direct simulations
of~\cite{BIMA} where $64\leq N \leq 1024$. This is in principle a
strong limitation for a model that is critical in the whole low
temperature phase with slowly decaying power law finite size
corrections.  For example for such small sizes the shape of the
probability distribution of the order parameter $P(q)$ is strongly
affected by finite size effects and is quite different from the
textbook shape of the infinite volume limit.

The results of Monthus and Garel for the SK model with Gaussian
couplings at temperature $T=T_c/2=0.5$ can be summarized as follows:
The disorder averaged logarithm of the largest relaxation time behaves
according to

\begin{equation}
\overline{\ln{\tau}} \opsim_{} \overline{B}/T \oppropto_{N\to\infty}
N^{\psi} \hskip 1cm \psi \simeq 0.33 \ ,
\end{equation}

on small systems ($N\leq 20$) already.
The probability density function of $\ln{\tau}$ scales like

\begin{equation}
P_N(\ln{\tau})= \frac{1}{\Delta} {\tilde P}
\Bigl(\frac{\ln\tau-\overline{\ln\tau}}
     {\Delta}\Bigr) \ ,
\label{PN}
\end{equation}

using the measured values of $\Delta^2\equiv\overline{(\ln\tau)^2}
-(\overline{\ln\tau})^2 $ and $\overline{\ln\tau}$, with an $N$
independent ${\tilde P}(.)$. Monthus and Garel are not able to measure
the width exponent $\psi_{width}$ defined by $\Delta\propto
N^{\psi_{width}}$ (which means that it is crucial to use the measured
values of $\Delta $ and $\overline{\ln\tau}$ in equation~\ref{PN}, at
least for small systems), but make a tentative indirect estimate
$\psi_{width}\simeq 0.26$ from their measurement of the tail exponent
$\eta$ (defined later in equation~\ref{tail}) and an assumption about
the disorder configurations that dominate the tail of ${\tilde P}(x)$
for $x\gg 1$.  In~\cite{JANKE} already, the quoted value of
$\psi_{width}\approx 0.25$ is lower than $1/3$.  We remark however
that the figure 1 of~\cite{JANKE} shows systematic errors as big as
$0.05$ in the value of $\psi$ from fits of data with $128 \leq N \leq
1024$, which are blamed on finite size effects. The agreement between
the results of~\cite{MONGAR} and~\cite{JANKE} for $\psi_{width}$ (that
is harder to measure than $\psi$) is accordingly somewhat surprising.

The purpose of this note is to investigate the distribution
$P_N(\ln{\tau})$ using the direct Monte Carlo simulation method, which
gives access to much larger system sizes than the method
of~\cite{MONGAR}, and without the strong assumptions about the
dynamics of the multi canonical algorithm made in~\cite{JANKE}.  We
take $1024$ disorder samples (with binary couplings) and measure for
each sample the dynamical overlap

\begin{equation}
q_J(t)=\frac{1}{N} \sum_{i=1}^N \sigma_i(t_0) \sigma_i(t+t_0) \ ,
\end{equation}

averaging over $t_0$ along a long trajectory starting from a well
equilibrated spin configuration~\footnote{These initial configurations
  have been obtained after $800000$ parallel tempering sweeps with
  $38$ temperatures uniformly distributed inside $[0.4,1.325]$. There
  are ample empirical evidences (see e.g.~\cite{ABMM}) that this is
  enough to achieve equilibrium, for values of $T\leq 0.4$ and $N$ up
  to $4096$, namely one obtains symmetric overlap probability
  distributions disorder sample by disorder sample, and the values of
  the internal energy and disordered averaged overlap squared
  $\overline{<q^2>}$ converge with excellent accuracy to the exact
  thermodynamic limit. Here $N\leq 512$ and there is in our opinion
  no doubt that equilibrium has been achieved.}, i.e. we measure the
thermal averaged $<q_J(t)>$ at equilibrium.  In practice a chain of
$10^8$ Metropolis sweeps (with sequential site update~\cite{SOK}) was
generated for two independent copies of the system (two clones),
starting from two independent spin configurations, for each disorder
sample. Measurements were made every $4$ sweeps (namely the average
was done over the values $t_0=0,4,8,\ldots$).  It would be a waste of
CPU time to measure the value of $q_J(t)$ for every (integer) values
of $t$.  It was measured for the following values: [1-20] with lag 1,
[22-40] with lag 2, [44-80] with lag 4, \ldots.  Altogether $q_J(t)$
was measured for 194 values of $t$, up to a maximum value of
$t_{max}\equiv 3670016$.  For $N=512$, a smaller chain of $4\ 10^7$
Metropolis sweeps was generated, and $q_J(t)$ was measured for 183
values of $t$ up to a maximum value of $t_{max}\equiv 1703936$ only.
The relaxation time $ \tau_J$ is defined by the condition
$q_J(\tau_J)=\sqrt{<q^2>_J}/2$, where the mean value of the static
overlap squared $<q^2>_J$ has been measured at equilibrium in the same
disorder sample.  Note that the ratio $q_J(t)/\sqrt{<q^2>_J}$ is
dimensionless and is accordingly~\footnote{In the critical regime, a
  spin-spin correlation function behaves like
  $\hat{q}(k,\omega,\xi)=\xi^{2-\eta+z}\ G(\xi k,\xi^z \omega)$ where
  $k$ is the momentum, $\omega$ the frequency, with some function
  $G(.)$.  Integrating over $k$ in order to have the $x=0$ correlation
  function and after Fourier transform, this equation becomes $
  q(x=0,t)\sim\xi^{-d+2-\eta} H(t/\xi^z)$ with some function
  $H(.)$. Using hyperscaling it mean that $ q(x=0,t)/\sqrt{<q^2>}$ is
  a function of $t/\xi^z\propto t/\tau$ only.} a function of
$t/\tau_J$.  Namely $q_J(t)/\sqrt{<q^2>_J}=F_J(t/\tau)$, with some
$F_J(.)$ that is a continuously decreasing function of its
argument. The precise analytical form of $F_J(.)$ is irrelevant.  We
have checked that, disorder sample by disorder sample, the difference
between $q_J(t)$ measured using clone one and $q_J(t)$ measured using
clone two is so small that both give in most cases indistinguishable
results for $ \tau_J$. (The worst case is for $N=64$ and $T=1$ with a
relative error of $0.07$ for a particular bad disorder sample, namely
the observed discrepancy is always smaller than $0.07$, when not
exactly zero, for more interesting values of $T$ and $N$.)  This shows
that the plus or minus one standard deviation estimates of $q_J(t)$
give the same estimate for $ \tau_J$ to an excellent precision, and
that accordingly the thermal noise is negligible.  That the thermal
errors are negligible can alternatively be seen as follows: At fixed
$J$, with $P$ independent measurements of $\ln\tau_J$, namely
$\ln\tau_J^{(1)},\ln\tau_J^{(2)},\ldots $, one has the elementary
unbiased estimator of $\delta_J$, the thermal statistical error on
$\ln\tau$, given by the expression $\delta_J^2\equiv
1/(P-1)\Bigl(1/P\sum_{i=1}^P (\ln\tau_J^{(i)})^2-(1/P\sum_{i=1}^P
\ln\tau_J^{(i)})^2 \Bigr) $.  In our simulation the two clones provide
two fully independent measurements of $\ln\tau_J$, and we have
accordingly the unbiased (but noisy) estimator
$\delta_J^2=1/4(\ln\tau_J^{(1)}-\ln\tau_J^{(2)})^2$. As mentioned
before, the ratio $\delta_J/\ln\tau_j$ is very small for all disorder
sample $J$ , system size $N$ and temperature $T$.  We note that the
disorder averaged $\sqrt{\overline {\delta_J^2}}$ is less than one
percent of the median of $\ln{\tau}$ for all values of $N$ and $T$
(For obvious reasons, we consider only the values of $N$ and $T$ such
that the median of ${\tau}$ is less than $t_{max}$, and the average of
$\delta_J^2$ is restricted to the disorder samples for which $\tau<
t_{max}$). The disorder average of ${ \sqrt{\delta_J^2}}$ is even
smaller.

The fact that the thermal noise is negligible is an essential
condition in order to obtain meaningful estimates for the probability
distribution of the relaxation time.  The various parameters (the
total run length, the window of measurement $t_{max}$, and the lag
between two successive measurements) have been chosen empirically,
they are such that the thermal noise is negligible (as we just said),
the CPU times spent in Monte Carlo updates and measurements are
balanced and the program fits inside the computer memory.  No attempt
was made to optimize theses parameters, it should in principle be done
separately for each disorder sample, each size $N$ and each
temperature $T$. We remark that the time scales considered
in~\cite{BIMA} are different from the one considered here,
indeed~\cite{BIMA} considers time scales (called $\tau_1$, $\tau_2$
and $\tau_3$ in~\cite{BIMA}) that can be defined from the time decay
of $q_J(t)$ measured with a single starting point $t_0$ on the one
hand, and time scales (called $\tau_q$ and $\tau_{q^2}$
in~\cite{BIMA}) defined from the time decay of the disorder averaged
$\overline{<q(t)>}$ (or $\overline{<q(t)^2>}$) on the other hand . The
time scales $\tau_1$, $\tau_2$ and $\tau_3$ have both thermal and
disorder fluctuations, whereas $\tau_q$ and $\tau_{q^2}$have no
disorder fluctuations by construction.  The time scales considered in
the present note are disorder dependent with negligible thermal noise
contamination. The purpose of~\cite{BIMA} was indeed to show that all
time scales in the SK model behave according to $\ln\tau \propto
N^{1/3}$.

\begin{figure}
\centering \includegraphics[width=0.3\textwidth,angle=270]{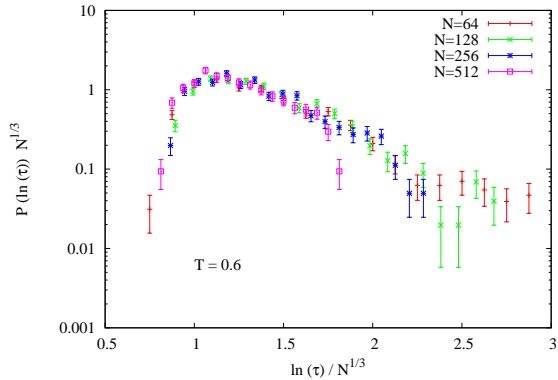}
\caption{(Color on line) Scaling of the probability distribution
  $P_N(\ln{\tau})$  at fixed temperature $T=0.6$.}
\label{figure1}
\end{figure}

We have data for $N=64$, $128$, $256$ and $512$, with temperatures
between $T=0.4$ and $T=1.1$ with steps of $0.1$ (the critical
temperature of the model is $T_c=1$). Figure~\ref{figure1} shows a
scaling plot of $P_N(\ln{\tau})N^{1/3}$ as a function of
$\ln\tau/N^{1/3}$ for $T=0.6$.  The relative statistical error on the
value of $P_N(.)$ inside a bin is $1/\sqrt{Q}$, with $Q$ the number of
data points inside the bin.  Figure~\ref{figure1} show good scaling,
confirming that $\overline{\ln{\tau}}$ has a behavior compatible with
$\psi=1/3$, namely (In this formula $\hat{P}(.)$ is an $N$ independent
function, related to the function $\tilde{P}(.)$ introduced above),

\begin{equation}
P_N(\ln{\tau})= \frac{1}{N^{1/3}} \hat{P}
\Bigl({\ln\tau}
     / {N^{1/3}}\Bigr) \ ,
 \label{master}
\end{equation}

extending the scaling of the probability distribution found
in~\cite{MONGAR} from $N\leq 20$ to $N\leq 512$, with both
$\overline{\ln \tau}$ and $\Delta$ explicitly proportional to
$N^{1/3}$.  Data taken at other values of the temperature (below
$T_c$) show similar scaling.  Depending on the temperature our
sampling of the tail of the probability distribution is limited by the
number of Monte Carlo sweeps performed (this is the case at low
temperature) or by the number of disorder samples used (this is the
case at higher temperature).  Since we are plotting the logarithm of
the (histogrammed) probability distribution of the logarithm of the
relaxation time, enlarging substantially the data range in
figure~\ref{figure1} would require a huge increase of the
computational effort.

\begin{figure}
\centering \includegraphics[width=0.3\textwidth,angle=270]{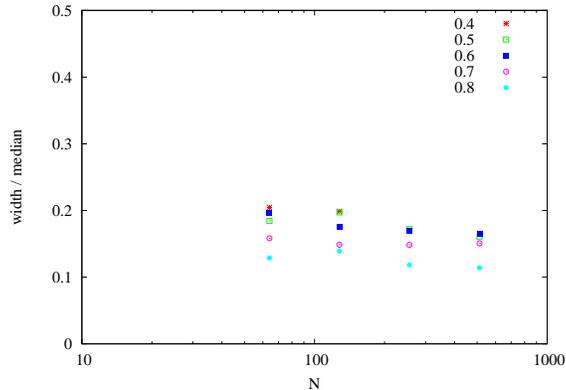}
\caption{(Color on line) The ratio of the width $W(N)$ of the
  distribution divided by the median $M(N)$, as a function of $N$ for
  several temperatures.}
\label{figure2}
\end{figure}

Implicit in figure~\ref{figure1} is that the width $\Delta$ scales
with the same exponent $1/3$ as the mean. This can be checked
directly: Following~\cite{BIMA} we have computed the median $M(N)$ of
the distribution of $\ln\tau$ and a width $W(N)$ defined arbitrarily
by $\int_{W(N)}^{M(N)} P_N(\ln{\tau}) \ d \ln {\tau} =0.30$.  The use
of the median of the distribution instead of the average, and of a
width defined from the quantiles of the distribution has the advantage
that the later is only defined if $\tau_J<t_{max}$ for all disorder
samples whereas the former requires that at least one half of the
disorder samples satisfy this bound.  The computational gain is
enormous with a distribution that has a very long tail towards large
values of $\tau$.  The ratio $W(N)/M(N)$ as a function of $N$ is shown
in figure~\ref{figure2} for several temperatures (without statistical
errors). It shows a weak $N$ dependence, consistent with the same
scaling for the width and the disorder average of the distribution. (As a
function of $T$ however the ration $W(N)/M(N)$ decreases slightly as $T$
grows.)  Indeed a plot of $P_N(\ln{\tau})W(N)$ as a function of
$(\ln\tau-M(N))/W(N)$ shows the same scaling as figure~\ref{figure1}.

Our conclusion for the behavior of the distribution of $\ln\tau$ with
$N$ is that, going to quite larger system sizes confirm the results of
Monthus and Garel.  With larger systems however, there is no need
anymore to use  the measured mean value and width of
the distribution, in scaling plots. Assuming
that both scale like $N^{1/3}$ just does it. Comparing to
previous~\cite{MACYOU,{COLBOR},BIMA} numerical determinations of the
exponent $\psi$, the fact that the whole distribution, and not only the
median, scales with the exponent $\psi=1/3$ strengthens the
conclusion that the exponent is indeed $\psi=1/3$.

The behavior of the function $\tilde P(x)$ for large values of $x$
defines the tail exponent $\eta$,

\begin{equation}
\ln {\tilde P(x)} \oppropto_{x\to\infty} -x^{\eta} \ ,
\label{tail}
\end{equation}

that is $\eta\simeq 1.36$ according to Monthus and
Garel~\cite{MONGAR}.  The value $\eta=1$ would imply a linear slope in
figure~\ref{figure1}, whereas the value $\eta= 1.36$ would imply a
slight downwards curvature. Both values are clearly compatible with
our data, that unfortunately do not sample deep enough inside the tail
of $P_N(\ln \tau)$ to allow a meaningful estimate of $\eta$.

\begin{figure}
\centering
\includegraphics[width=0.3\textwidth,angle=270]{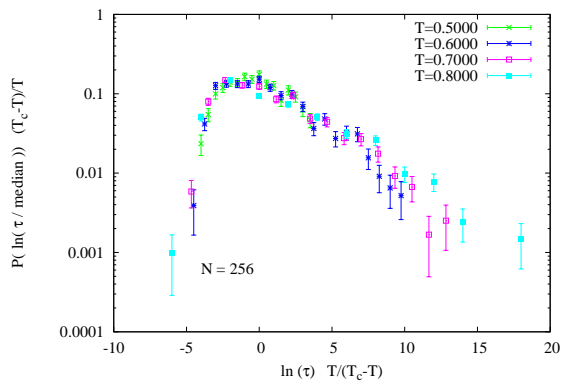}
\caption{(Color on line) Scaling at fixed size $N$ of the probability
  distribution $P(\ln{\tau}-\overline{\ln{\tau}} )$. Here $N=256$.}
\label{figure4}
\end{figure}

Since we have data for several values of the temperature, we can see
if our probability distributions scale with the temperature also. We
have seen that, at fixed temperature, the distribution of the largest
barrier scales like $N^{1/3}$. Since the Sherrington-Kirkpatrick model
is a mean field model, the scaling combination is $N^{1/(\nu
  d_{up})}(T_c-T)$, where $d_{up}=6$ is the upper critical dimension
of the theory and $\nu=1/2$~\cite{Brezin1, Brezin2}.  The scaling of
the probability distribution of the relaxation time has been known for
some time for the spherical Sherrington-Kirkpatrick, and is of this
form indeed~\cite{RODMOO},

\begin{equation}
\ln{\tau_J}\sim \frac{N^{1/3}}{T} (T_c-T) {\bf R}\ ,
\end{equation}
with ${\bf R}$ a random variable (independent of $N$ and $T$).  Here
$N^{-2/3} \  {\bf R}$ is the difference between the two largest
eigenvalues of the coupling matrix of the model.  Mathematical proof of this
scaling behavior and the expression for the probability distribution
of ${\bf R}$ can be found in~\cite{TW}.

In figure~\ref{figure4} we check the hypothesis that indeed
$\ln\tau_J=\overline{\ln{\tau}} +(T_c-T)N^{1/3} {\bf R}/ T$ with ${\bf
  R}$ a random variable (with a distribution that does not depend on
$N$ or $T$). This figure shows reasonable scaling, even if the
temperatures are not so close to $T_c$. Indeed the scaling quality
deteriorates if one adds points closer to $T_c$. This seems
paradoxical, the likely explanation is that $\tau_J$ is not a pure
exponential and that sub-leading power law contributions to $\tau_J$
becomes important close to $T_c$ where the dominant term in the
exponential goes to zero, since at $T_c$ one has $\tau\propto
N^{z/d_{up}}$.  In order to obtain a good scaling at fixed $N$ it is
crucial to consider the distribution of
$\ln\tau_J-\overline{\ln{\tau}} $ as is done in figure~\ref{figure4}.
There is no such need at fixed $T$ like in figure~\ref{figure1}. The
likely explanation is that $\overline{\ln{\tau}} $ behaves like
$N^{1/3}$ but not like $(T_c-T)/T$, or that sub-leading power law
contributions to $\tau_J$ are important.

\begin{figure}
\centering
\includegraphics[width=0.3\textwidth,angle=270]{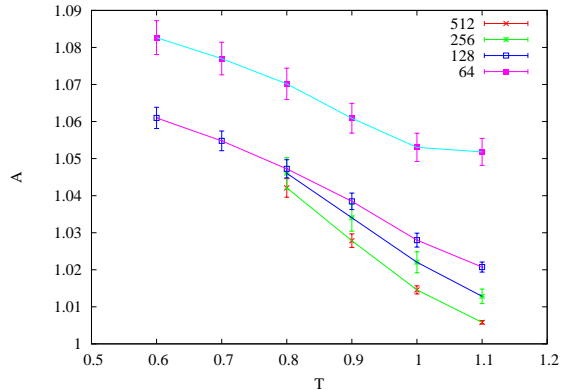}
\caption{(Color on line). The ratio
$A=\overline{(\ln{\tau})^2}/\overline{\ln{\tau}}^2$ as a function of the
  temperature, for $N=64$ to $512$.}
\label{figure5}
\end{figure}

Figure~\ref{figure5} shows the ratio
$A=\overline{(\ln{\tau})^2}/\overline{\ln{\tau}}^2$, where the overline
is here the usual arithmetic average (using the median would give a trivial
result). It should be a constant, independent of $N$ and $T$,
if $\tau_J$ was proportional to $(T_c-T)N^{1/3}$. Small
scaling violations are visible, the data for increasing number of
spins $N$ seem to converge toward a value with a small temperature
dependence.  This is in agreement with our remark about the behavior
of $\overline{\ln{\tau}} $ as function of $N$ and $T$ in the paragraph
above.  We remark en passant that if following~\cite{MONGAR}
and~\cite{JANKE} we had $\psi_{width}<\psi$, then $A=1+C
N^{\psi_{with}-\psi}$,with some constant $C$, and should accordingly
converge towards one, with unfortunately extremely slowly decaying
corrections.

The quality of our data does not allow a
trustable determination of the Kurtosis of the distribution
$G={\overline{(\ln{\tau}-\overline{\ln\tau})^4}/}
{\Bigl[\overline{(\ln{\tau}-\overline{\ln\tau})^2}\Bigr]^2}$. For
$N=256$ and $512$
the value of the Kurtosis is strongly affected by a few
rare disorder samples with large values of $\ln\tau$. Restriction the
analysis to $N=64$ and $128$, one obtains values of $G \approx 5$ with
little or no temperature dependence, but finite size effects.

In conclusion, we have measured the probability distribution of the
equilibrium relaxation time of the Sherrington-Kirkpatrick model with
binary couplings, for a range of system size and temperature. We
checked that our estimates are free of thermal noise. The data confirm
that both the average and the width of the probability distribution of
$\ln\tau$ scale as $N ^{1/3}$ in the spin glass phase.  As a
last remark, we note that the long tail of the distribution of $\tau$
may be a hidden source of severe problems in numerical simulations
with the widespread practice of using the same number of Monte Carlo
iterations for all disorder samples. This has been notices several
time already, but is may be worth repeating.

\section{Acknowledgments}  This work 
originates in a remark made by Jean-Philippe Bouchaud. I thanks him
and Giulio Biroli, Bertrand Eynard, Thomas Garel, Alexandre Lef\`evre
and C\'ecile Monthus for enlightening discussions. The numerical
simulation were done using the CCRT computer center in
Bruy\`eres-le-Ch\^atel.

\end{document}